\documentclass[reprint,prl,aps,twocolumn,showpacs,reprint,superscriptaddress,nofootinbib,citeautoscript]{revtex4-1}
\usepackage{amsmath,amssymb,hyperref}
\usepackage{graphicx}
\usepackage{datetime}
\usepackage{soul}
\usepackage{braket}
\usepackage{xfrac}
\usepackage{upgreek}
\usepackage[usenames,dvipsnames]{color}
\usepackage{natbib}
\usepackage{hyperref}
\usepackage{appendix}
\usepackage{titlesec}
\usepackage{ulem}
\usepackage{nameref}
\begin{document}  
\title{Learning the Fuzzy Phases of Small Photonic Condensates} 
\author{Jo\~{a}o D. Rodrigues}
\email[Corresponding author:]{j.marques-rodrigues@imperial.ac.uk}
\affiliation{Physics Department, Blackett Laboratory, Imperial College London, Prince Consort Road, SW7 2AZ, United Kingdom}
\author{Himadri S. Dhar}
\affiliation{Physics Department, Blackett Laboratory, Imperial College London, Prince Consort Road, SW7 2AZ, United Kingdom}
\author{Benjamin T. Walker}
\affiliation{Physics Department, Blackett Laboratory, Imperial College London, Prince Consort Road, SW7 2AZ, United Kingdom}
\affiliation{Centre for Doctoral Training in Controlled Quantum Dynamics, Imperial College London, Prince Consort Road, SW7 2AZ, UK}
\author{Jason M. Smith}
\affiliation{Department of Materials, University of Oxford, Oxford, UK}
\author{Rupert F. Oulton}
\affiliation{Physics Department, Blackett Laboratory, Imperial College London, Prince Consort Road, SW7 2AZ, United Kingdom}
\author{Florian Mintert}
\affiliation{Physics Department, Blackett Laboratory, Imperial College London, Prince Consort Road, SW7 2AZ, United Kingdom}
\author{Robert A. Nyman}
\affiliation{Physics Department, Blackett Laboratory, Imperial College London, Prince Consort Road, SW7 2AZ, United Kingdom}
\begin{abstract}
\begin{footnotesize}
Phase transitions, being the ultimate manifestation of collective behaviour, are typically features of many-particle systems only. Here, we describe the experimental observation of collective behaviour in small photonic condensates made up of only a few photons. Moreover, a wide range of both equilibrium and non-equilibrium regimes, including Bose-Einstein condensation or laser-like emission are identified. However, the small photon number and the presence of large relative fluctuations places major difficulties in identifying different phases and phase transitions. We overcome this limitation by employing unsupervised learning and fuzzy clustering algorithms to systematically construct the fuzzy phase diagram of our small photonic condensate. Our results thus demonstrate the rich and complex phase structure of even small collections of photons, making them an ideal platform to investigate equilibrium and non-equilibrium physics at the few particle level.
\end{footnotesize}
\end{abstract}
\maketitle
%
%
\par
Phase transitions are extraordinary manifestations of collective behaviour that mark abrupt changes in the properties of many-particle systems. The associated discontinuities in the thermodynamic quantities~\cite{Menon1995, Chu2012, Ensher1996, Lipa2003}, which allow an unequivocal identification of both phases and phases transitions, can only emerge in the limit of infinite degrees of freedom~\cite{Yang1952, Huang1987} however. Intuitively, extensive quantities $Q$, like energy or particle number, show fluctuations of the order $\sqrt{Q}$, with $\sqrt{Q}/Q$ vanishing in the thermodynamic limit $Q\rightarrow \infty$, thus giving rise to sharp transitions in large systems. Photonic condensates have emerged as powerful platforms for exploring the fundamental physics of phase transitions and critical phenomena. Equilibriun Bose-Einstein condensation of light has been achieved in diverse platforms, including dye-filled microcavities~\cite{Klaers2010, Nyman2015, Greveling2018}, plasmonic lattices~\cite{Hakala2018} or fibre cavities~\cite{Weill2019}. Dye-filled microcavities are particularly interesting, as driving, loss and thermalization rates can be independently controlled to give access to rich non-equilibrium regimes~\cite{Kirton2013, schmitt2015thermalization, Hesten2018, Walker2018, Walker2020}. Further, the ability to precisely engineer the trapping potential~\cite{Dung2017, Kurtscheid2019} results in an impressive degree of control over these systems, including the capacity to generate ensembles of only a few photons~\cite{Walker2018}, begging the question of whether or not collective behaviour and different phases of matter can still be identified.
\begin{figure}
\centering
\includegraphics[scale=0.80]{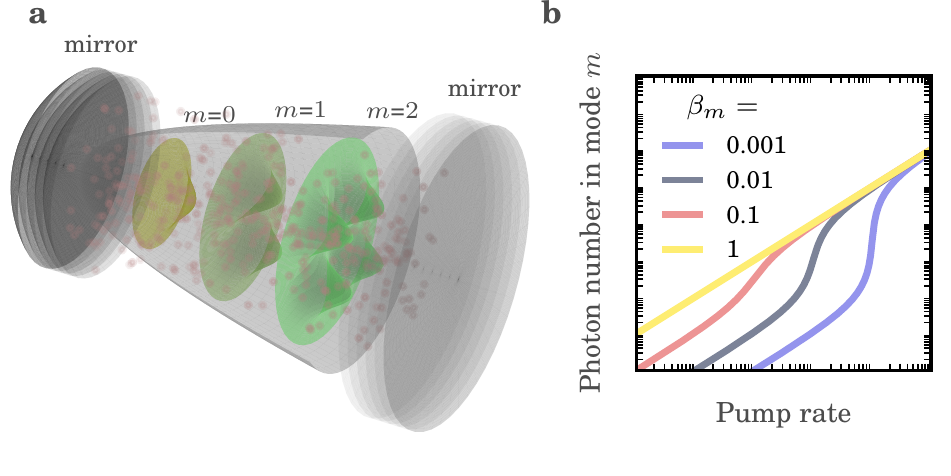}
\caption{Small photonic condensates: a, Schematic of our multi-mode dye-filled microcavity. The small radius of curvature induces a tight harmonic potential, whose eigenmodes are depicted in colors. The resulting large mode spacing highly suppresses the thermal excitation of highly-excited modes. Due to transverse radial symmetry, non-degenerate photonic modes are labeled by a single quantum number $m$. Their spatially-dependent profiles and the resulting mode-mode competition can bring the system into complex non-equilibrium states exhibiting collective behaviour even at low photon numbers. b, Illustration of the effect of the system size, quantified by the parameters $\beta_m$, on the criticality of the lasing phase transition, in the limit of negligible mode-mode coupling. Small systems, $\beta_m \rightarrow 1$, are shown to exhibit broad transitions, while pure criticality is recovered in thermodynamic limit, $\beta_m^{-1} \rightarrow \infty$.}
\label{fig:scheme}
\end{figure}
\begin{figure*}
\centering
\includegraphics[scale=0.80]{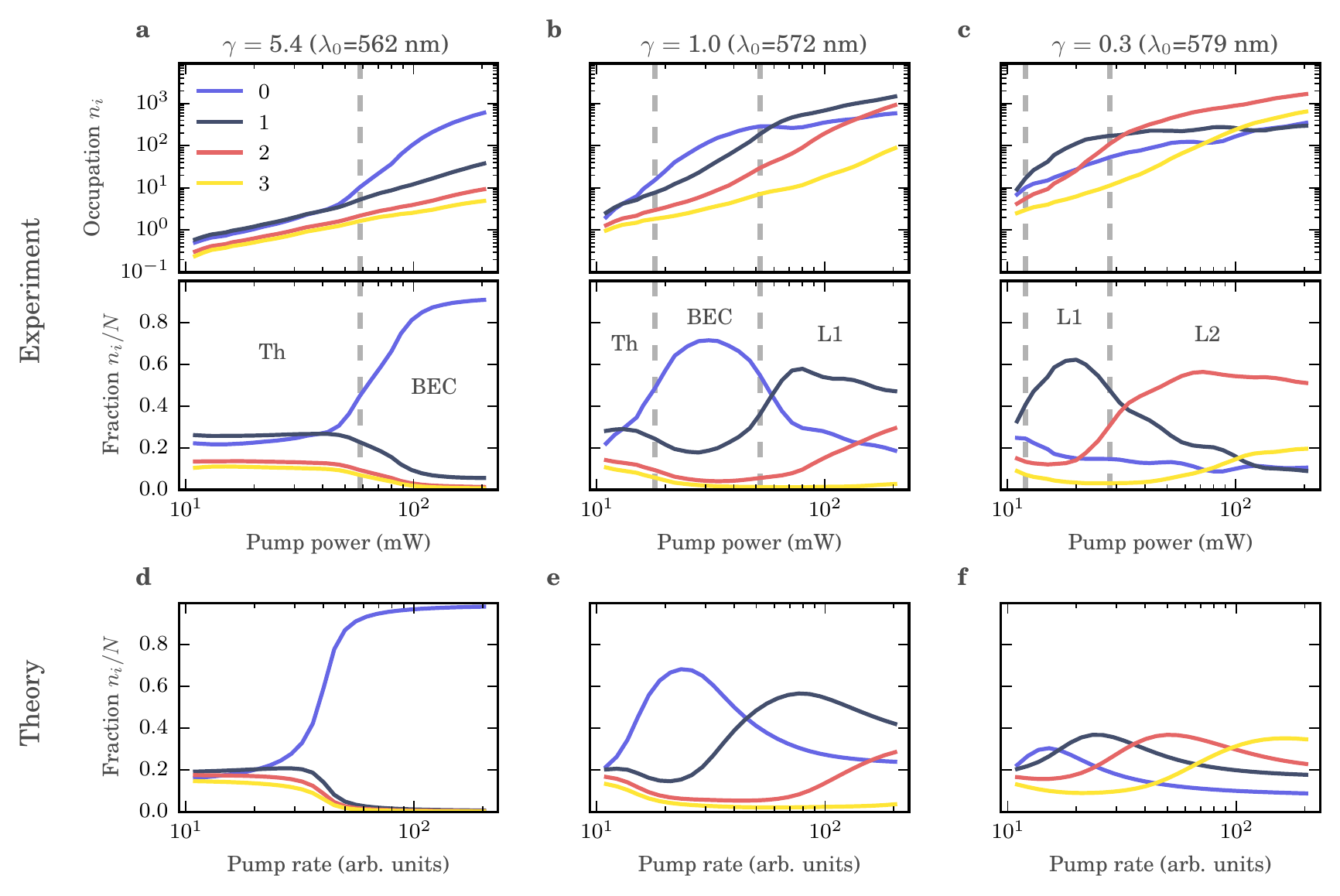}
\caption{Occupation numbers: a-c, Absolute and relative occupations numbers of the first four photonic modes, for different thermalization regimes. The latter is experimentally tuned by sweeping the cavity cutoff wavelength $\lambda_0$. Each set of occupation numbers defines a configuration. The dashed vertical lines mark transitions between different representative phases, which are labeled as: Th: thermal phase; BEC: Bose-Einstein condensate; L1: condensed first excited-state; L2: condensed second-excited state. d-f, Results from the non-equilibrium model of photonic condensation.}
\label{fig:populations}
\end{figure*}
\par
Other physical systems are known where, despite the macroscopic number of particles, the identification of thermodynamic phases and phase transitions can still be hindered, for example, by high-dimensional configuration spaces, or through the existence of nontrivial order parameters, like in topological~\cite{Huse2013, Zhang2017} or many-body localized states~\cite{vanNieuwenburg2017,Schindler2017}. Machine learning techniques, such as neural networks, have been shown to successfully detect and classify such complex phases of matter, mostly due to their ability to retrieve the often few significant features in otherwise large sets of data. These frameworks, however, require prior knowledge of the phase structure of the system's Hamiltonian, falling into the domain of supervised learning~\cite{Carleo2017, Chng2017, Venderley2018, Dong2019, Pilozzi2018, Pilozzi2021}, with few experimental results yet reported. Some examples include the training of a neural network far from the critical region and the posterior characterization of the Mott insulator-superfluid transition~\cite{Rem2019}, or the usage of an artificially synthesized dataset, carefully designed to reflect the expected symmetries of a nematic phase in electronic quantum matter~\cite{Zhang2019}. Unsupervised learning, on the other hand, has received less attention in the physical sciences. Examples include the study of simple spin systems~\cite{Wang2016} or, more recently, the detection of topological phase transitions~\cite{kaming2021}. Unrelated with the detection of phases and phase transitions, unsupervised learning and neural networks can also be used as generative models, in the context of quantum-state reconstruction ~\cite{Torlai2018, Torlai2019}.
\par
Here, we explore the phases of a small photonic condensate in a dye-filled microcavity. The trapping potential is engineered to combine a small cavity volume with a large mode spacing, as shown in Fig.~(\ref{fig:scheme}). The parameter $\beta_{m}$ is defined as the fraction of spontaneous emission into the $m^\mathrm{th}$ cavity mode, generalizing the standard $\beta$ parameter introduced in the context of single-mode microlasers~\cite{Rice1994} into the realm of multi-mode systems. In the absence of mode-mode coupling and photon re-absorption, the lasing, or condensation, transition occurs at a photon number that scales as $\beta_m^{-1}$, while the corresponding threshold width is of the order $\sqrt{\beta_m}$. Criticality is thus recovered in the large cavity limit, i.e. the thermodynamic limit $\beta_m^{-1} \rightarrow \infty$. Our multi-mode microcavity operates in a mesoscopic regime characterized by the existence of collective behaviour despite the small system size. Such a mesoscopic regime, however, being inherently characterized by large relative fluctuations of order $\sqrt{Q} /Q$, strongly inhibits our ability to detect and characterize different phases and phase transitions. Details of the experiment, the underlying physics of microlasers and photonic condensation in microcavities, as well as a theoretical model describing these processes can be found in the Supplementary Material~\cite{sup}. 
\begin{figure*}
\centering
\includegraphics[scale=0.80]{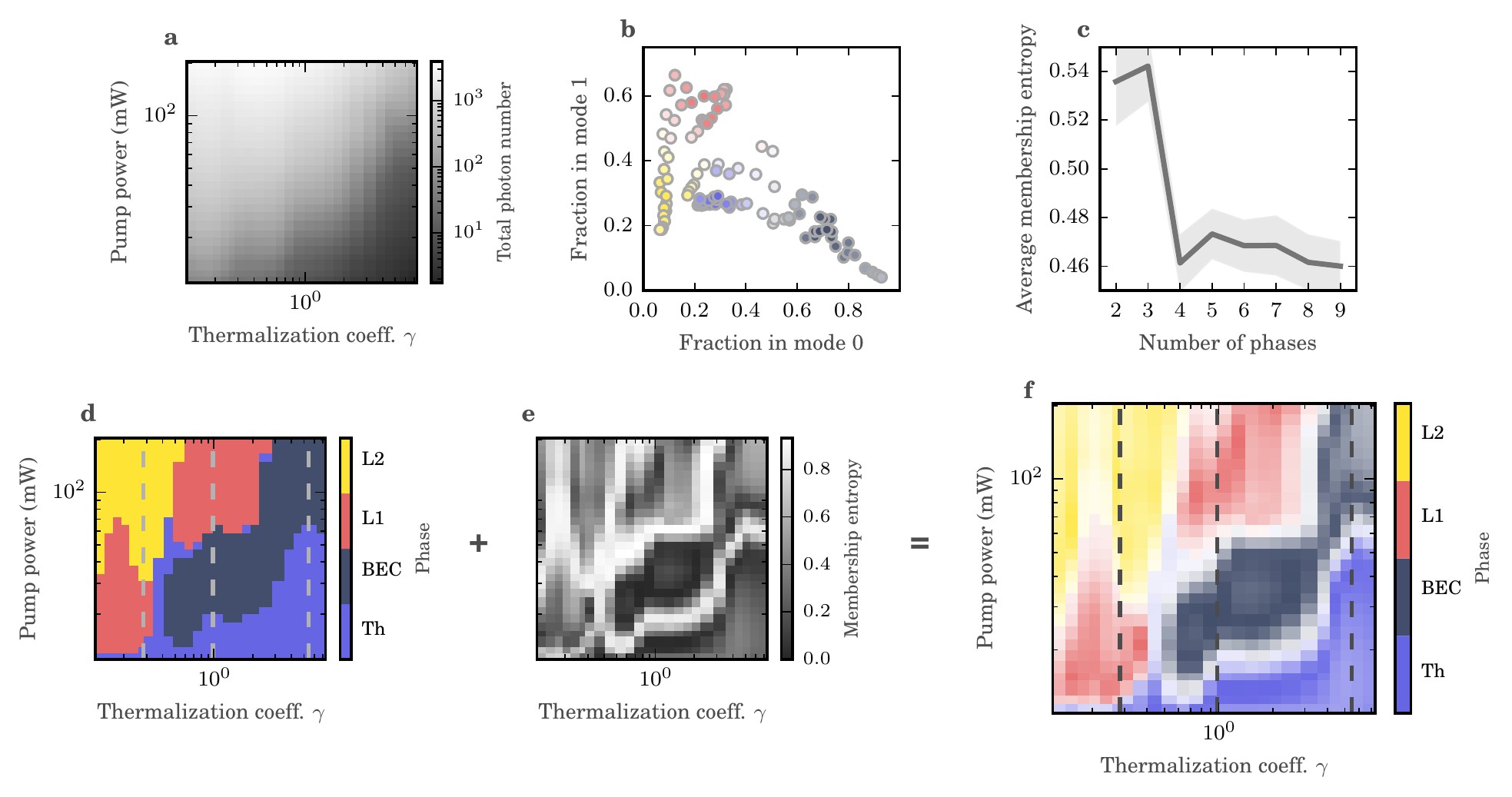}
\caption{Fuzzy phase diagram. a, Total photon number as a function of pump power and thermalization coefficient. b, Projection of the full ten-dimensional feature space onto the two-dimensional plane spanned by the relative occupation of the ground and the first excited state, after the fuzzy clustering procedure being applied. Different representative phases are shown by different colors, with a transparency level proportional to the respective membership entropy. c, Estimation of the number of phases by minimizing the average membership entropy, whose distribution is approximated by bootstrapping the feature space. The resulting standard deviation is depicted by the dashed grey area. This ensures the stability of this procedure. d, Representative phase diagram. e, Membership entropy. The fuzzy phase diagram in (f) is constructed from the representative phase diagram in (d) by assigning a transparency level proportional to the membership entropy in (e). In this way, regions of larger phase ambiguity becoming white, thus visually gauging the degree of representativeness of the representative phase diagram. The dashed vertical lines indicate the traces depicted in Fig.~(\ref{fig:populations}).}
\label{fig:phase_diagram}
\end{figure*}
\par
The photonic modes inside the microcavity are those of a two-dimensional harmonic oscillator, each with a degeneracy proportional to the mode number $m$. There is a single $m$=0 ground-state located at approximately 540~THz or, equivalently, 560~nm. Excited states are separated by roughly 2.1~THz, which is only slightly smaller than the thermal energy scales, with only a few low-energy modes becoming thermally accessible~\cite{sup}. Photon thermalization results from multiple emission and absorption events with the dye molecules. These occur at rates $E_m$ and $A_m$, respectively, which are related by the Kennard-Stepanov relation $A_m = E_m e^{-\delta_m/k_BT}$, with the detuning $\delta_m = \omega_{\mathrm{ZPL}}-\omega_m$ from the zero-phonon line $\omega_{\mathrm{ZPL}}$ of the dye molecules. We work on the Stokes side of the molecular transition, where $E_m > A_m$. Constant pumping of dye excitations is required to maintain steady-state operation due to the finite mirror transmission, which is quantified by the cavity loss rate $\kappa$. Since the finite cavity lifetime limits the thermalization process, we define the thermalization coefficient $\gamma=A_0/\kappa$ as the average number of absorption events per cavity lifetime, with $A_0$ the absorption rate at the cavity cutoff. Regimes of good thermal contact with the molecular reservoir imply fast thermalization, i.e. $\gamma \gg 1$. 
\par
Besides the driven-dissipative character described above, additional processes occurring in the microcavity contribute to the emergence of complex non-equilibrium behaviour, as schematically depicted in panel ({\bf a}) of Fig.~(\ref{fig:scheme}). The heterogeneity among the different mode functions gives rise to a spatially-dependent competition for the finite molecular excitations. The result is a form of incoherent mode-mode coupling mediated by dye molecules, an effect which becomes more noticeable at higher pump powers and is responsible for the breaking of thermal equilibrium. 
\par
By tuning the thermalization coefficient and the pump power, we can thus bring our photonic system into distinct equilibrium and non-equilibrium regimes, as shown in Fig.~(\ref{fig:populations}). Regions of good thermal contact are characterized by the existence of a wide thermal regime at low pump power followed by a smooth transition into a state where most photons occupy the ground-state alone, consistent with a Bose-Einstein condensate. On the contrary, under weaker thermal contact excited states are shown to become highly populated as the pump power is increased, indicating a breakdown of thermal equilibrium. A non-equilibrium model derived from a full quantum description accurately describes these observations, particularly in the regimes of strong thermalization~\cite{sup}. For weaker thermalization, the mode condensation dynamics are highly dependent on imperfections in the shape and alignment of the pump beam, which is at the origin of the slight deviations between theory and experiment depicted in Fig.~(\ref{fig:populations}).
\par
Despite the qualitative observations above, systematically inferring the existence or not of different phases and phase transitions remains ilusive. Certainly, clear manifestations of any form of criticality are mostly absent. This is neither a peculiarity of the system at study nor a limitation of the measurement apparatus, but rather a fundamental statistical consequence of the small particle number and the complete breakdown of the thermodynamic limit. Here, we overcome this limitation by turning to a machine-augmented approach and, as a result, introduce the idea of fuzzy phases. Importantly, methods known so far use supervised learning techniques, which require the prior knowledge of the Hamiltonian phase structure, used to train models capable of inferring the phase of unlabelled configurations~\cite{Carrasquilla2017, Chng2017, Venderley2018, Dong2019, Rem2019, Zhang2019}. In contrast, here we start completely devoid of any knowledge about our photonic system, thus necessarily falling into the domain of unsupervised learning, which will allow us to recover the subtle phase structure of our photonic condensate in a data-driven and model-free approach.
\begin{figure}
\centering
\includegraphics[scale=0.88]{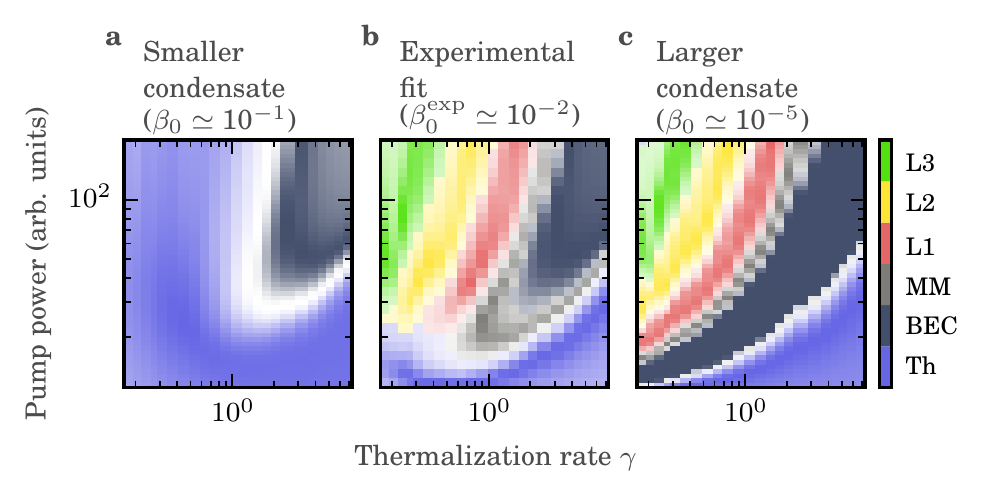}
\caption{
Fuzzy and sharp photonic condensates: b, Fuzzy phase diagram obtained from the non-equilibrium model of photonic condensation, with parameters that match the experiment. As described earlier, the model slightly deviates from the experiment at weak thermalization conditions, where it predicts the condensation of the third excited cavity mode, the L3 phase. At intermediate thermalization rates, we also infer the presence of a multi-mode (MM) phase, characterized by the condensation of both the ground and the first excited state. This is, however, a relatively narrow and fuzzy phase which is not resolved in the experiment. The theoretical model is also used to generate configurations for both smaller (a) and larger (c) photonic condensates, quantified by the fraction of spontaneous emission into the ground-state. The thermodynamic limit is approached from left to right.}
\label{fig:small_vs_large}
\end{figure}
\par
We consider the relative occupation numbers of the ten lowest-energy photonic modes. A configuration becomes a point in this bounded ten-dimensional feature space and different configurations are spanned by changing the pump power and thermalization coefficient, these last two defining the parameter space. The search for structure proceeds with the clustering of nearby points in feature space, under a similarity metric, here taken as the simple Euclidean distance. This is schematically depicted in panel (\textbf{b}) of Fig.~(\ref{fig:phase_diagram}). Each cluster contains similar configurations which are maximally distinguishable from those of the remaining clusters, being thus identifiable with a particular phase. We point out that it is important to uniformly sample the parameter space, such that all inferred structure becomes uniquely linked to the intrinsic phase properties of the system at study. The smooth transitions between phases suggests the use of fuzzy logic, where the association of configurations and phases becomes probabilistic. More precisely, for each configuration, i.e., each point $x$ in the feature~space, we wish to find the probabilities $p_x(i)$, with $i=1,2,...,k$ and $k$ the estimated number of phases, such that $p_x(i)$ is the probability of configuration $x$ belonging to the $i^\mathrm{th}$ phase, thus gauging a level of membership between phases and configurations. The {\it fc-means} algorithm~\cite{Dunn1973, Bezdek2013} computes exactly these membership probabilities by minimizing the overall weighted distance between points in feature space and the cluster centroids -- check Supplementary Material for mode technical details on the fuzzy clustering procedure implemented here~\cite{sup}.
\par
Such a probabilistic model allows us to quantify the level of ambiguity in associating configurations with phases and introduce the idea of fuzzy phases. Here, the fact that physical systems made up of only a few particles often exhibit properties simultaneously consistent with multiple phases, becomes inherent to the whole formalism, and is precisely quantified by the set of membership probabilities $p_x(i)$. From these, we can define the representative phase which, for a fixed configuration $x$, is taken as the phase $i$ with the highest membership probability $p_x(i)$, if it exists. We also define the membership entropy as
\begin{equation}\label{eq:entropy}
S_x \left( \lbrace p_x(i) \rbrace \right) = - \frac{1}{\mathrm{log}(k)} \displaystyle{\sum_i} p_x(i) \mathrm{log}(p_x(i)),
\end{equation}
thus quantifying the fuzziness of a given phase. It is normalized such that a maximally fuzzy configuration, or maximally fuzzy phase, has unit entropy, corresponding to $p_x(i) = 1/k$. On the other hand, definite phases are recovered in the limit $S_x=0$, corresponding to $p_x(j)=1$ and $p_x(i\neq j)=0$.
\par
The learned phase structure of our small photonic system is depicted in Fig.~(\ref{fig:phase_diagram}). Four different photonic phases are estimated by minimizing the average membership entropy, depicted in panel (\textbf{c}). We begin by constructing the representative phase diagram, shown in panel (\textbf{d}). By inspection of the bosonic occupation numbers in Fig.~(\ref{fig:populations}), the different phases can be associated with: a thermal phase (Th); a Bose-Einstein condensate (BEC); and the condensation of the first (L1) and second (L2) excited states. Note that, while instructive, the information contained in the representative phase diagram alone is fundamentally incomplete. The membership entropy further complements the picture and, together with the representative phase diagram, allows us to construct the fuzzy phase diagram, depicted panel (\textbf{f}), which directly reflects the fundamental absence of critical behaviour inherent to the few-particles regime. Here, instead, different representative phases are separated by broad regions of large membership entropy or, equivalently, large phase ambiguity. Supplementary Fig. (\textbf{1}) further demonstrates this behaviour.
\par
The idea of fuzzy phases can be further explored by considering the non-equilibrium theoretical model of photonic condensation, as depicted in Fig.~(\ref{fig:small_vs_large}). Here, we simulate systems of different size by changing the parameter $\beta_m$. We begin by simulating a smaller system than that of the experiment, $\beta_m > \beta_m^\mathrm{exp}$. Here, the membership entropy (fuzziness) increases across the entire phase diagram. As a result, the optimal classification essentially retains the existence of only the thermal and the Bose-Einstein condensed phase. Previously inferred phases in larger systems are now blurred together with the thermal phase, as it no longer becomes relevant to consider them as independent phases. This is not arbitrarily imposed but rather optimally inferred by the fuzzy clustering procedure acting on observational data alone. On the other hand, for a larger system, $\beta_m < \beta_m^\mathrm{exp}$, the membership entropy across the phase diagram becomes smaller and mostly concentrated in the increasingly narrower regions between representative phases. In the limit of infinite number of particles (thermodynamic limit), the membership entropy vanishes everywhere, as each configuration is uniquely associated with a particular phase with unit probability, except in the infinitesimally narrow regions regions marking pure phase transitions.
\par
The increasing ability to control and prepare small atomic~\cite{Serwane2011, Murmann2015} and photonic~\cite{Bohnet2012, Dung2017, Walker2018} systems has given rise to a great deal of scientific interest in few-particle physics~\cite{Schiulaz2018, Tomasz2019}. In this work, we have demonstrated that small collections of photons in dye-filled microcavities can be prepared in a wide range of both equilibrium and non-equilibrium configurations, ranging from thermal equilibrium distributions and Bose-Einstein condensates, to non-equilibrium laser emission. The systematic inference of such rich phase structure, however, was shown to be severely limited by the presence of large relative fluctuation and the resulting absence of critical behaviour. This limitation was overcome by employing machine learning techniques, where criteria for defining phases and phase transitions are neither phenomenologically or axiomatically imposed, but rather optimally inferred from observational data alone. While the clustering approach is applicable for systems of any size, the probabilistic nature of the fuzzy logic algorithms becomes particularly relevant in small system. Importantly, the fuzzy character does not reflect a state of incomplete knowledge but rather a fundamental statistical implication of the small particle number. We anticipate immediate applications of these techniques in the investigation of how collective effects emerge from a bottom-up approach, as the system's size is gradually increased~\cite{Wenz2013}, or in the investigation of how magnetic phases change in the few-particle limit~\cite{Murmann2015_2}. A distinct, yet exciting possibility, would be the identification of fuzzy phases in the context of liquid phase condensation inside biological cells, where the formation of membrane-less coherent structures seems to depend on smoothly varying concentration thresholds, suggesting the presence of significant finite-size effects~\cite{Shineaaf2017, Riback2020}.
\nocite{*}
%
%
%
%
\par
We acknowledge financial support from EPSRC (UK) through the grants EP/S000755/1 and EP/N010868/1, the Centre for Doctoral Training in Controlled Quantum Dynamics EP/L016524/1 and the European Commission via the PhoQuS project (H2020-FETFLAG-2018-03) number 820392. We also thank Aur\'{e}lien Trichet for technical assistance fabricating the mirror substrates and Rick Mukherjee for discussions on machine learning algorithms.
%
\bibliography{references}
%
%
\end{document}